\documentclass[default]{archive_submission}

\usepackage{lineno} 
\usepackage[]{rotating}
\usepackage{float}

\jyear{2022}%

\theoremstyle{thmstyleone}%
\theoremstyle{thmstyletwo}%
\usepackage{comment}
\theoremstyle{thmstylethree}%
\begin{document}


\title[\,]{Portrait of locally driven quantum phase transition cascades in a molecular monolayer}


\author*[1,2,3]{\fnm{Soroush} \sur{Arabi}}\email{soroush.arabi@fkf.mpg.de}

\author[2,4]{\fnm{Taner} \sur{Esat}}

\author[2,4,5]{\fnm{Aizhan} \sur{Sabitova}}

\author[2,3]{\fnm{Yuqi} \sur{Wang}}

\author[6]{\fnm{Hovan} \sur{Lee}}

\author[6]{\fnm{C\'{e}dric} \sur{Weber}}

\author[3,7]{\fnm{Klaus} \sur{Kern}}

\author[2,4,5]{\fnm{F.~Stefan} \sur{Tautz}}

\author*[2,8]{\fnm{Ruslan} \sur{Temirov}}
\email{r.temirov@fz-juelich.de}

\author*[1,2,4]{\fnm{Markus} \sur{Ternes}}\email{ternes@physik.rwth-aachen.de}

\affil[1]{\orgdiv{Institute of Physics IIB}, \orgname{RWTH Aachen University}, \orgaddress{52074 Aachen}, \country{Germany}}

\affil[2]{\orgdiv{Peter-Gr\"{u}nberg-Institute for Quantum Nanoscience}, \orgname{Research Center J\"ulich}, \orgaddress{\city{52425 J\"ulich}, \country{Germany}}}

\affil[3]{Max Planck Institute for Solid State Research, Heisenbergstraße 1, 70569 Stuttgart, Germany}
\affil[4]{J\"ulich Aachen Research Alliance, Fundamentals of Future Information Technology, 52425 J\"ulich, Germany}

\affil[5]{\orgdiv{Institute of Physics IV}, \orgname{RWTH Aachen University}, \orgaddress{\city{52074 Aachen},  \country{Germany}}}

\affil[6]{King's College London, Theory and Simulation of Condensed Matter (TSCM), The Strand, London WC2R 2LS, UK}

\affil[7]{Institut de Physique, \'{E}cole Polytechnique F\'{e}d\'{e}rale de Lausanne, 1015 Lausanne, Switzerland}

\affil[8]{Institute of Physics II, University of Cologne, 50937 Cologne, Germany}

\abstract{
Strongly interacting electrons in layered materials give rise to a plethora of emergent phenomena, such as unconventional superconductivity \cite{cao2018(1),Yankowitz2019,Chen2019}, 
heavy fermions \cite{Vano2021, Yazdani2016, Hunt2013}, and spin textures with non-trivial topology \cite{sharpe2019,Liu2022,Du2009}. Similar effects can also be observed in bulk materials, but the advantage of two dimensional (2D) systems is the combination of local accessibility by microscopic techniques and tuneability. In stacks of 2D materials, for example, the twist angle can be employed to tune their properties \cite{Novoselov2016, Andrei2021}. However, while material choice and twist angle are global parameters, the full complexity and potential of such correlated 2D electronic lattices will only reveal itself when tuning their parameters becomes possible on the level of individual lattice sites.
Here, we discover a lattice of strongly correlated electrons in a perfectly ordered 2D supramolecular network by driving this system through a cascade of quantum phase transitions
using a movable atomically sharp electrostatic gate. 
As the gate field is increased, the molecular building blocks
change from a Kondo-screened to a paramagnetic phase one-by-one, enabling us to reconstruct their complex interactions in detail. 
We anticipate that the supramolecular nature of the system will in future allow to engineer quantum correlations in arbitrary patterned structures \cite{Khajetoorians2019, Gomes2012, Drost2017, Green2014}.
}

\maketitle

The prototypical building block in strongly correlated systems is a localised magnetic moment immersed in a bath of itinerant electrons. 
Below a characteristic temperature, the Kondo temperature $T_{\rm K}$, the hybridisation between the local moment and the itinerant electrons forms a many-body singlet state at the Fermi energy $E_{\rm F}$, the so-called Kondo singlet, and results in the quenching of the local moment. 
Arranging such building blocks periodically leads to the lattice version of this phenomenon, i.\,e., a Kondo lattice.
Despite their complexities, these systems are typically understood in terms of the competition of two antagonistic effects:
The local antiferromagnetic Kondo interaction which screens the local moments at every site of the lattice by the spin of the conduction electrons, and a long-range interaction between different local moments that favours magnetic ordering and breaking the Kondo singlets \cite{Doniach1977}. 
In a subtle balance of these two effects, close to absolute zero temperature, the systems become critical, and quantum fluctuations dominate the physics \cite{Coleman2005, Lohneysen2007,Kirchner2020}.

Whereas the understanding of the Kondo lattice physics has been developed to a high degree 
through the study of transition-metal and heavy-fermion crystals, attempts to access the local information regarding the quantum degrees of freedom (spins) and to coherently manipulate them \cite{Heinrich2021, ternes2017} are still intensively stimulating  the field. 
Naturally, this striving has redirected the approach to realising  strongly correlated lattices from three-dimensional stoichiometric compounds \cite{Aynajian2010, Kavai2021} towards artificial spin structures in reduced dimensions and also promoted the utilisation of local probes for measurements. 
In this regard, studies have shown the formation of  correlated lattices using moir\'{e} superlattices in 2D transition metal dichalcogenides \cite{Vano2021,Kumar2022}, and alternatively, in bottom-up approaches using molecular/atomic building blocks which have been arranged either by atomic manipulation techniques \cite{Figgins2019, Moro-Lagares2019} or by self-assembly   \cite{Jiang2011,Ternes2004}.  
Yet, despite the enormous progress, a detailed microscopic understanding, particularly from the local real-space perspective, is still out of reach. 

\begin{figure}[t]
    \centering
    \includegraphics[width=0.99\textwidth]{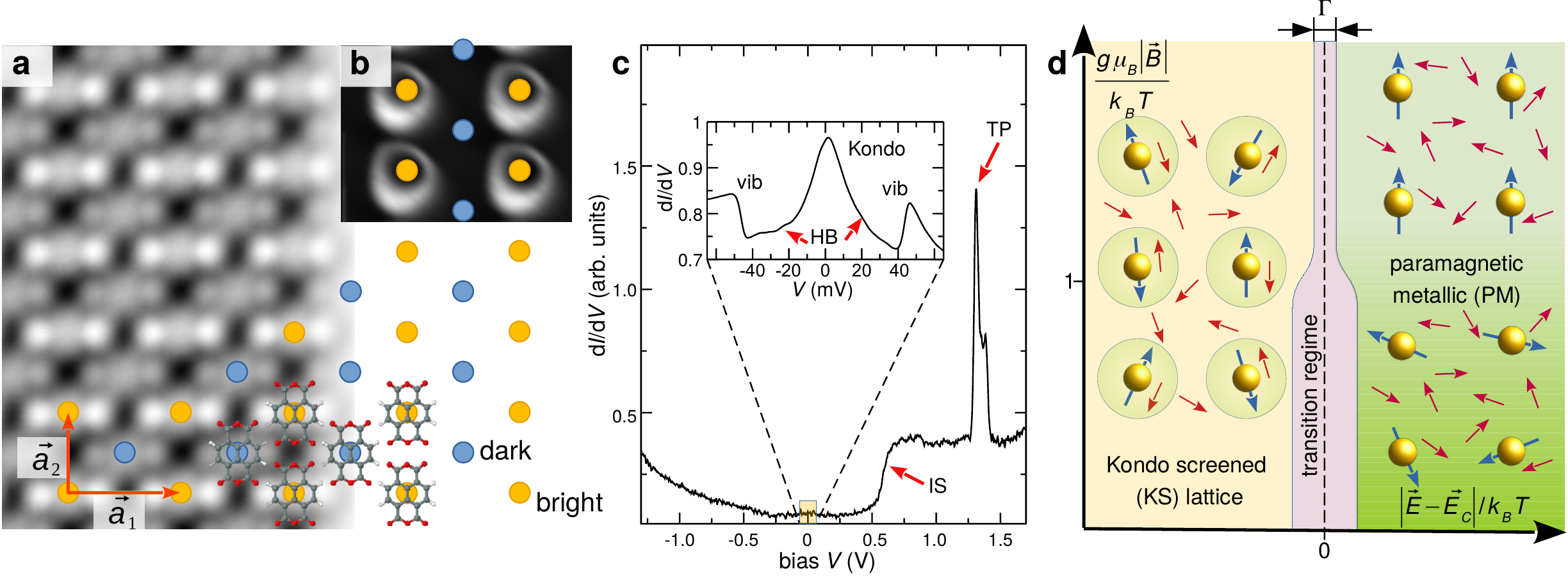}
    \caption{\textbf{The correlated molecular lattice:} \textbf{a,} Constant-current gray-scale STM image ($V=-48\,\mbox{mV}, I = 1\,\mbox{nA}$) of a monolayer of NTCDA molecules on Ag(111) showing a brick wall like structure consisting of molecules with higher (brighter) and lower (darker) appearance (yellow and blue  dots, respectively). The apparent height difference is $\approx 30\,\mbox{pm}$. 
    Superimposed are stick-and-ball models of the molecules and the lattice basis vectors $\mathbf{a_{1}}$ and $\mathbf{a_{2}}$ with lengths $1.5$ and $1.16\,\mbox{nm}$. \textbf{b,} Constant-current gray-scale differential conductance ($dI/dV$) image measured at higher bias ($V=1.4\,\mbox{V}, I=6\,\mbox{nA}, V_{\rm mod}=8\,\mbox{mV}$) revealing ring-like structures centred at each bright molecule.
    \textbf{c,} $dI/dV$ spectrum (setpoint: $V=2\,\mbox{V}, I=4\,\mbox{nA}, V_{\rm mod}=8$\,meV) measured with the tip of the STM positioned above the centre of a bright molecule shows two prominent features: the interface state (IS) of the Ag(111) surface, and a sharp transition peak (TP) which is at the origin of the ring-like structures seen in panel (b). 
    The inset shows an over the lattice averaged $dI/dV$ spectra close to the Fermi energy $E_{\rm F}$. It reveals molecular vibration modes (vib), a Kondo resonance, and Hubbard bands (HB) introduced by the lattice. \textbf{d,} Schematic phase diagram of the lattice versus applied electric ($\vec{E}$) and magnetic ($\vec{B}$) field scaled by temperature $T$ (for $T\ll T_{\rm K}$ and $\mu_{\rm B}\lvert \vec{B} \rvert \ll k_{\rm B}T_{\rm K}$). The Kondo screened phase (KS) exists at $\vec{E}<\vec{E}_{\rm C}$ and undergoes a quantum phase transition to a paramagnetic metallic (PM) phase constituting free spins. The width 
    $\Gamma$ of the transition regime where fluctuations dominate the system narrows with the $\vec{B}$-field. A quantum critical point at $T=0$ is expected to exist where the phases merge.}
    \label{fig: fig1}
\end{figure}

Therefore, to fill this gap, we study a self-assembled relaxed monolayer of 1,4,5,6-naphthalene tetracarboxylic acid dianhydride (NTCDA) on an Ag(111) substrate \cite{Stahl1998, Fink1999}. 
In gas phase the molecule is charge neutral, but upon adsorption to Ag(111) its (formerly) lowest unoccupied molecular orbital shifts below $E_{\rm F}$ ($E_{\rm SOMO}\approx\!-0.5$\,eV) and becomes a singly occupied molecular orbital (SOMO) with spin 1/2 \cite{Bendounan2007, Schoell2010, Tonner2016, Jakob2016}.
In low-temperature scanning tunnelling microscopy (STM), the molecules are imaged at low bias as dumbbell-shaped objects forming a brick wall lattice with alternating rows of topographically higher (brighter) and lower (darker) appearance  (Fig.\,\ref{fig: fig1}a) \cite{Eickhoff2019}.
The superlattice unit cell contains, therefore, two slightly different molecules: a bright one and a dark one.

Interestingly, imaging the same lattice at large positive bias $V >1.3$\,V reveals ring-like structures in the differential conductance $(dI/dV)$ maps, centred on the bright molecules (Fig.\,\ref{fig: fig1}b).
These rings originate from a strong transition peak (TP) which can also be seen in the single-point spectroscopy above the bright molecules (Fig.\,\ref{fig: fig1}c).
In the following, we will show that the TP indicates a quantum phase transition (QPT) in the molecular layer in which the electric field of the STM tip drives the Kondo-screened molecules of the lattice successively into a paramagnetic free spin phase (Fig.\,\ref{fig: fig1}d). 

We focus now on energies close to $E_{\rm F}$ where we find very complex and strongly position-dependent spectra in the lattice (see Supplementary Note 1). 
A spectrum spatially averaged over the lattice  (Fig.\,\ref{fig: fig1}c inset) reveals a broad peak close to zero bias as a manifestation of Kondo singlets.
Furthermore, the spectrum contains conductance steps at $\approx\pm\,47$\,meV, which are due to inelastic vibrational excitations of the individual molecules \cite{Braatz2012,Tonner2016, Eickhoff2019}. 
At the flanks of the Kondo peak, we see well-resolved electron- and hole-like quasiparticle excitations, which will turn out to be Hubbard bands (HB).

\begin{figure}[t]
    \includegraphics[width=0.99\textwidth]{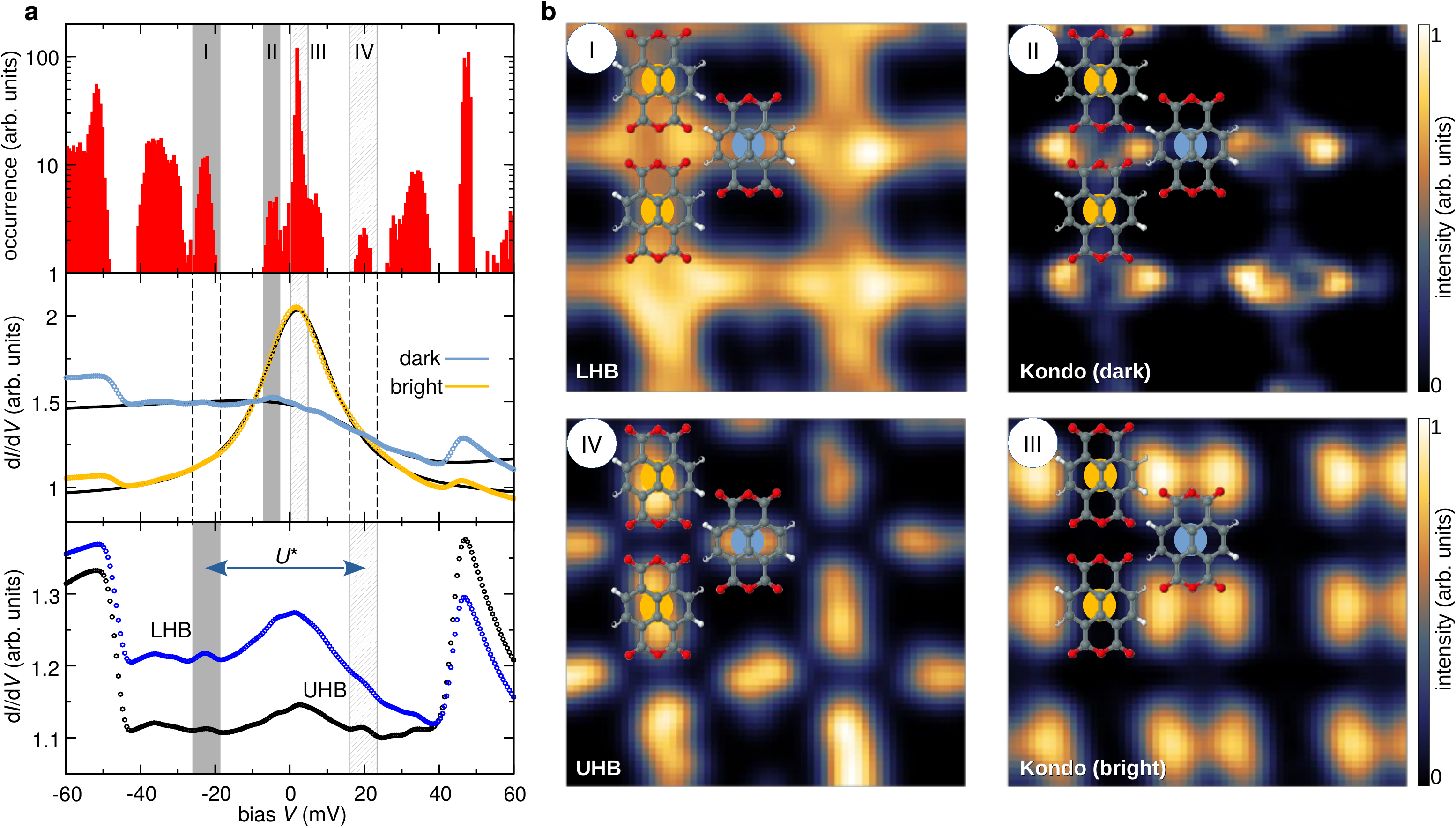}
    \caption{\textbf{Low energy features of the lattice. a,} Top: Energy distribution histogram of peak occurrences in a map of $58\times58$ $dI/dV$ spectra (setpoint for each spectrum: $V=70\,\mbox{mV}, I=1\,\mbox{nA}, V_{\rm mod}=0.2\,\mbox{mV}$) taken on the lattice. The full (hatched) bars mark areas at negative (positive) bias ranges where spectra have been averaged or which are mapped (panel b). Middle: Averaged Kondo resonance measured at bright and dark molecules and corresponding Fano fits (full lines) ($\Gamma =12.2\,(28.7)\,\mbox{meV}, q>300\,(=1.2), \epsilon_0=2\,(21)\,\mbox{meV}, T_{\rm K}=132\pm 5\,(340\pm 20)\,\mbox{K}$; values in brackets belong to dark molecules). Bottom: Averaged spectra of the LHB (blue dots) and UHB (black dots). \textbf{b,} Color-coded maps of the peak feature intensity in the energy ranges as marked in panel (a) and superimposed stick-and-ball model of the molecules.}
    \label{fig: fig2}
\end{figure}

To disentangle the spectroscopic features, we analyse individual spectra taken on a grid of $58\times58$ points on the molecular lattice (see Methods). 
In this dataset, we search for peak structures and plot their weighted occurrence in an energy distribution histogram 
(Fig.\,\ref{fig: fig2}a). 
This histogram shows two peaks very close to $E_{\rm F}$ whose averaged spectra correspond to the Kondo resonances of the bright and dark molecules, apparent in the occurrence maps (Fig.\,\ref{fig: fig2}b II and III). 
Compared to the bright molecules' $T_{\rm K}=132\pm5$\,K, the averaged spectrum of the dark molecules' Kondo resonance reveals, in agreement with ref.\,\cite{Eickhoff2019}, a higher $T_{\rm K}=340\pm 20$\,K due to a stronger hybridisation with the Ag(111). 
It shows instead of a peak an asymmetric Fano line-shape which is the result of interference between two possible tunnelling channels in the STM junction, i.\,e., electrons can tunnel from the tip either via the resonance or directly into the substrate \cite{Fano1961, Ternes2009}.
The Kondo resonances are strongly localised at the ends of the dumbbell-like topographic shape of the molecules, which correspond to their C--H sites \cite{Eickhoff2019}. At the centre of the molecule, however, the resonance vanishes. Such distribution cannot originate from a symmetric orbital, but points to a $\pi$-like molecular orbital with a nodal plane along the molecular long axis \cite{Eickhoff2019}.
This $\pi$-like orbital contains the spatially distributed spin with a phase shift $\theta=\pi$ between the two lobes and enables a significant intermolecular coupling \cite{Esat2016}.

Also, the states in the flanks of the Kondo peak are clearly distinguishable in the histogram with the maximum occurrence at $-23$\,mV and $+20$\,mV. Due to their low intensity, their averaged spectra are outshone by the vibrational excitations and the Kondo resonance.
Their mapping (see Methods), however, clearly reveals that the occupied state at negative bias is most strongly located between the molecules, while the unoccupied one is localised at each dark and bright molecule's centre 
(Fig.\,\ref{fig: fig2}b\,I and IV). 
This distribution, which is reminiscent of bonding and antibonding combinations between the $\pi$ states on the three neighbouring molecules indicated in Fig.\,\ref{fig: fig2}b, reveals that the flank peaks are a consequence of the formation of a correlated  lattice and hence can be naturally understood as lower and upper Hubbard bands (LHB/UHB)  \cite{coleman_2015}.
They have a composite structure containing contributions from the localised $\pi$-electrons of the molecules as well as from the conduction electrons. Note that their $U^\ast\approx 43$\,meV is much smaller than the Coulomb repulsion $U$ of the SOMO. The additional bands energetically further away from $E_{\rm F}$ also show spatial dependence (see Supplementary Note 2), but are for the following not important.

\begin{figure}[t]
    \includegraphics[width=0.99\textwidth]{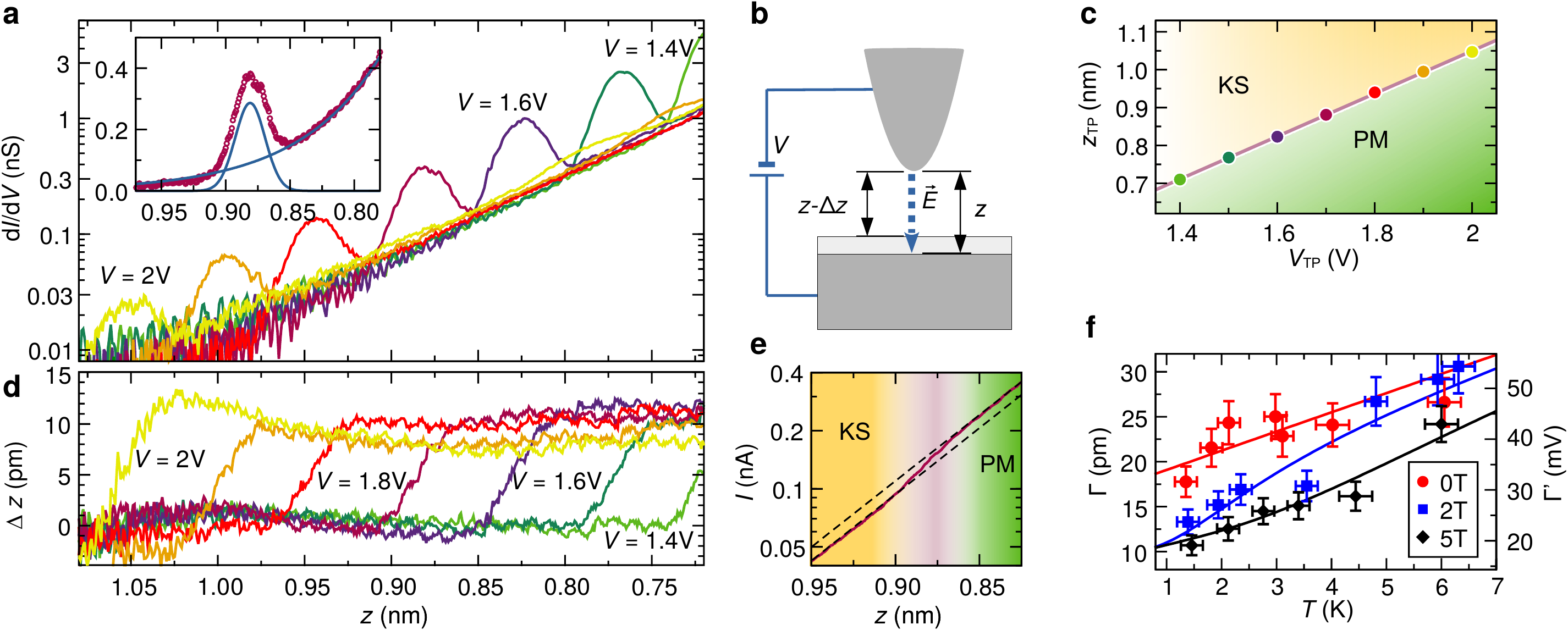}
    \caption{\textbf{Signatures of the quantum phase transition.} \textbf{a,} Logarithmic plot of the $dI/dV(z)$ signal above a bright molecule at constant $V$ ranging from $1.4 - 2.0$\,V and $B_z = 5$\,T. The peaks mark the QPT 
    between the KS 
    and the 
    PM state.
    The inset shows exemplarily the $V = 1.7\,$V data in a linear plot and a least-squares fit to the sum of an exponential and a Gaussian.  
    \textbf{b,} Model of the tip-sample tunnelling barrier $z$ and its change to $z+\Delta z$. \textbf{c,} Extracted Gauss centre $z_{\rm TP}$ from (a) versus $V_{\rm TP}$. 
    \textbf{d,} Extracted $\Delta z$ 
    from simultaneously measured $I(z)$ data in (a). 
    \textbf{e,} Logarithmic plot of the $I(z)$ signal for $V = 1.7\,$V with exponential fits (dashed lines) separating the two regimes of the system. 
    \textbf{f,} Half-width at half-maximum $\Gamma$ of the phase transition at different $\vec{B}$-fields and $T$ (dots) measured at $V=1.7V$, and least-square model ($R^2=0.91$) of a QPT to a paramagnetic spin 1/2 (solid lines). 
    }
    \label{fig: fig3}
\end{figure}

We now turn to the prominent TP feature of Fig.\,\ref{fig: fig1}c by recording $dI/dV(z)$ spectra above the centre of the bright molecules by changing the tip-sample tunnelling distance $z$ at successively reduced constant bias $V$ (Fig.\,\ref{fig: fig3}a). 
We define $z=0$ as the extrapolated height at which quantum of conductance would be reached.
The $z_{\rm TP}$ at which the peak occurs is proportional to $V_{\rm TP}$  (Fig.\,\ref{fig: fig3}c), leading us to an electric field driven origin of the TP.
Simplifying the STM junction as a plate capacitor with an electrical field between tip apex and sample of $\vec{E}=(V/z)\vec{e}_z$, with $\vec{e}_z$ as the unit vector along the $z$-direction (Fig.\,\ref{fig: fig3}b), the linear relation 
corresponds to a critical field $\lvert\vec{E}_{\rm C}\rvert=(1.77\pm 0.02)$\,GVm$^{-1}$ at which the TP occurs.

The TP in $dI/dV$ is equivalent to an increase in conductance, as the example of a simultaneously recorded $I(z)$ trace in Fig.\,\ref{fig: fig3}e illustrates. Until the TP is reached the current increases exponentially with $I\propto \exp(-\kappa z)$ and a $\kappa=15.3$\,nm$^{-1}$. 
Past the TP, $I(z)$ increases further with identical $\kappa$, excluding a modification of the tunnel barrier by any significant charge accumulation on the molecule. 
Thus, the TP is tantamount to a change of the effective tip-sample distance from $z$ to $z-\Delta z$. 
This $\Delta z\approx 10$\,pm is bias independent (Fig.\,\ref{fig: fig3}d), meaning that its origin lies in an increase of the molecular lattice conductivity.

Gating of a molecular charge state by exerting the $\vec{E}$-field of an STM tip has been observed \cite{Fernandez_Torrente2012, Pham2019, Kumar2019, Mohammed2020, Li2021, Yan2021} and is usually explained in the framework of a double barrier tunnel junction model in which the electric field shifts a single particle molecular state across $E_{\rm F}$, thereby changing the number of electrons on the molecule. 
This is different here. 
The only charge-like (and therefore $\vec{E}$ gateable) state which lies close to $E_{\rm F}$ is the LHB which is a composite state of the lattice with relatively weak intensity. 
The prominent Kondo resonance is fixed to $E_{\rm F}$ and, therefore, cannot be gated.

The width $\Gamma$ of the TP shows a peculiar behaviour: With increasing temperature $T$, $\Gamma$ increases linearly as expected from the thermal broadening of the Fermi edge of the tip (Fig.\,\ref{fig: fig3}f red line).
However, an additionally applied $\vec{B}$-field perpendicular to the surface of strength 
$g\mu_{\rm B}\lvert \vec{B}\rvert>k_{\rm B}T$ (with $\mu_{\rm B}$ as Bohr's magneton, and $k_{\rm B}$ as Boltzmann's constant) decreases $\Gamma$ significantly, as shown in Fig.\,\ref{fig: fig3}f. 
This behaviour is reminiscent of the freezing out of spin-fluctuations in an ideal paramagnetic metal with $g=2$ and can be well described by assuming a broadening due to the sum of intrinsic quantum and thermal fluctuations:
\begin{linenomath*}
\begin{equation}
    \Gamma=\Gamma_{\rm inst}+\Gamma_{0}\left(\overline{n}+\gamma\frac{k_{\rm B}T}{E_{\rm C}} \sqrt{\overline{n}}\right),
    \label{eq:gamma}
\end{equation}
\end{linenomath*}
with $\Gamma_{\rm inst}=1.6\pm1.3$\,pm as the finite resolution of our experimental setup and $\Gamma_{0}=7.7\,\pm 0.9$\,pm as the extrapolated quantum fluctuations per (spin) degree of freedom at $T=0$ where all thermal fluctuations are frozen out. 
The average 
number of available spin states, $\overline{n}=\exp(S/k_{\rm B})$ is given by the entropy $S$ of the paramagnetic spin 1/2 in the field $\vec{B}$ and at temperature $T$. $\gamma=4.0\,\pm\,1.3$ is the slope of the $T$-dependent broadening of the TP. 
Note, however, that this model neglects any magnetic interactions between neighbouring spins. Significant magnetic ordering would reduce $\overline{n}$ by freezing out fluctuations similar to the $\vec{B}$-field here (for further details see Methods).

The proportionality between $z_{\rm TP}$ and $V_{\rm TP}$ allows us to express $\Gamma_{0}$ also as a width in $V$, $\Gamma_{0}'=\Gamma_{0}\times \lvert \vec{E}_{\rm C} \rvert=14\pm2$\,mV (see Supplementary Note 3). From the applied bias $V$ only a small fraction $\alpha$ shifts the LHB (see Supplementary Note 4), which we can estimate for $V_{\rm TP}=1.3$\,V as $\alpha\approx E_{\rm LHB}/eV_{\rm TP}\approx 1/55$. This allows to estimate the width of the quantum fluctuation per degree of freedom in energy as $\Delta_{0}=\alpha\times\Gamma^{\prime}_{0} \approx 0.3$\,meV, and with this the timescale of the quantum fluctuation $\tau_{0}=\hbar/\Delta_{0}\approx 5$\,ps (with $\hbar$ as the reduced Planck's constant) of the QPT extrapolated for $T=0$.

\begin{figure}[t]
    \includegraphics[width=0.99\textwidth]{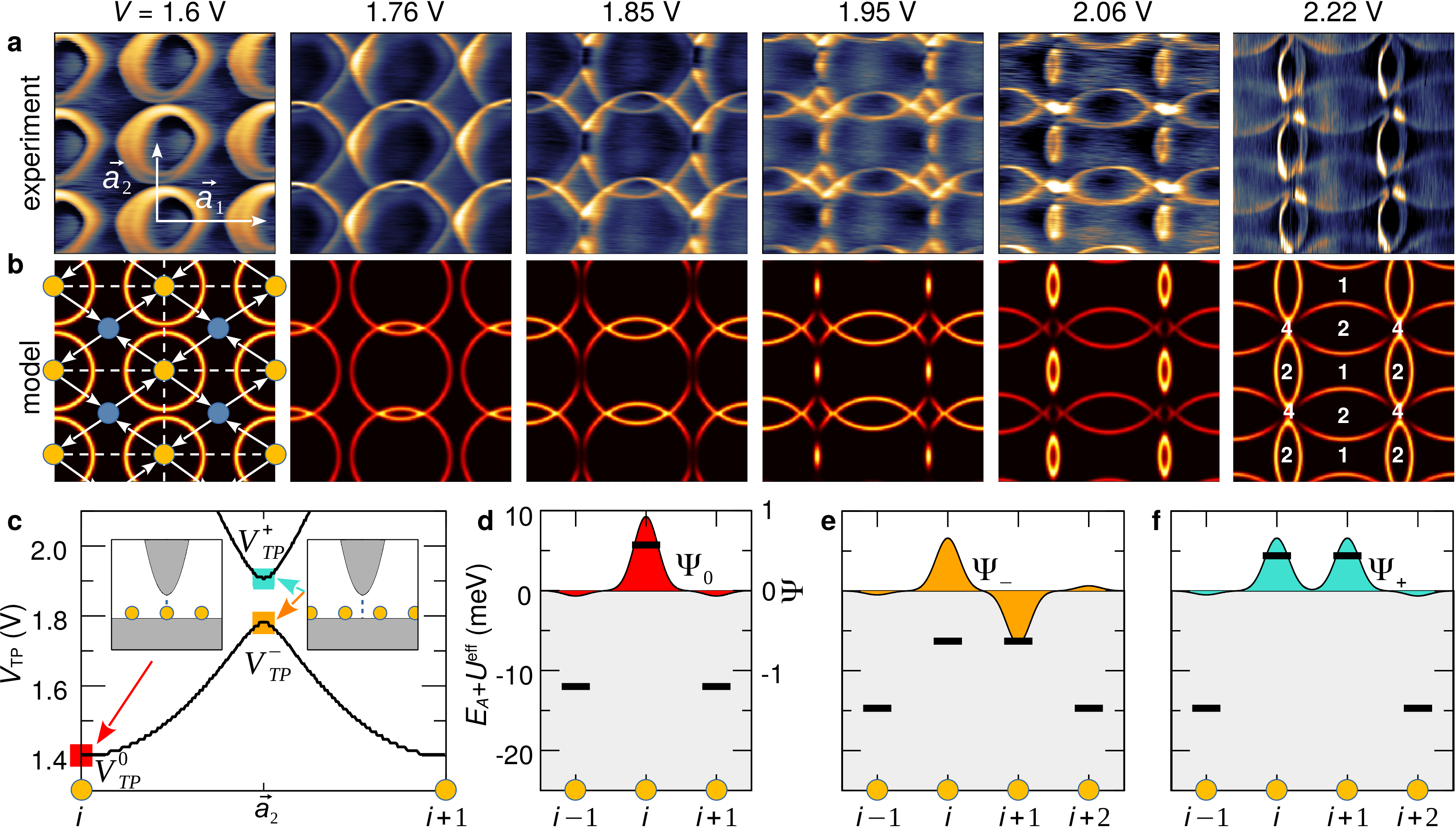}
    \caption{ \textbf{A portrait of strongly correlated quasiparticle excitations. a,} Colour-coded $dI/dV$ plots measured in an area of approx.\,$1.5\times1.5\,$nm$^{2}$ with a bright molecule in the centre  ($I=4.9$\,nA, $B_{z}=10$\,T). 
    \textbf{b,} TB simulation using a nearest-neighbour hybridisation $V_{\rm AB}e^{i\theta}$, with $V_{\rm AB}=4$\,meV and $\theta=0.2\pi$ between bright and dark molecules (arrows; their direction indicates hopping with positive phase shift), a next-nearest neighbour hoping $t_{\rm A}=0.25(0.05)$\,$\mu$eV between bright molecules in direction $\vec{a}_1$($\vec{a}_2$) (dashed lines), $t_{\rm B}=0$, and on-site energies (yellow: bright $E_{\rm A}=-23$\,meV, blue: dark $E_{\rm B}=20$\,meV). For the tip we used $r=16$\,nm and $\beta=2$. 
    The numbers in the last sub-panel indicate the number of QPTs in the lattice. \textbf{c,} Calculated $V_{\rm TP}$ for a tip at different positions along $\vec{a}_{1}$. Insets illustrate the tip position above or in between two bright molecules for the TPs occurring at $V_{\rm TP}^{0,+,-}$ whose wavefunctions $\Psi_{0,+,-}$ (coloured curves) and on-site energies $E_{\rm A}+U^{\rm eff}$ (black bars) of neighbouring sites are shown in panel \textbf{d-f}, respectively.}
    \label{fig: fig4}
\end{figure}

This QPT is driven by the $\vec{E}$-field of the tip. For a radially symmetric tip, the field decreases with the lateral distance $d=\sqrt{x^2+y^2}$ to the tip apex as $\vec{E}(d)= \vec{E}_{d=0} /\left(1+(d/r)^{\beta}\right)$, whereby $r$ and $\beta$ approximate the decay of the $\vec{E}$-field due to the shape of the tip \cite{Battisti2017}. 
At biases and tip heights where $\vec{E}_{d=0}>\vec{E}_{\rm C}$ the TP occurs at a certain lateral distance from the molecular centre, leading to growing ring-like structures when $dI/dV$ images are taken at successively increased $V$.
Fig.\,\ref{fig: fig4}a shows a set of such images. 
At $V=1.6$\,V similar rings as in Fig.\,\ref{fig: fig1}b form. 
The small deviation from perfect circularity can be assigned to some tip shape asymmetries. 
When rings from neighbouring bright molecules cross along the short $\mathbf{a}_2$ direction, they overlap. 
Contrarily, when they cross along the long $\mathbf{a}_1$ direction a gap opens (at $V=1.85$\,V), signalling an avoided crossing due to  intermolecular interaction \cite{Li2021}. 

To rationalise the ring formation in our bipartite lattice we use a single-particle tight-binding model in which we include the effect of the $\vec{E}$-field of the tip. 
We use a complex hybridisation amplitude $V_{\rm AB}e^{i\theta}$ between nearest-neighbour bright (A) and dark (B) molecules and different next-nearest-neighbour hopping between bright molecules along the directions $\vec{a}_1$ and $\vec{a}_2$ as illustrated in Fig.\,\ref{fig: fig4}b, left panel (for details see Methods).

The use of a finite phase $\theta$ is inspired by the seminal work of Haldane on graphene, in which he assumes a phase-dependent hopping between next-nearest neighbours \cite{Haldane1988}, which is based on the observation that electrons face different energy landscapes on their left and right while hopping forward and backward \cite{Kane2005}.
Similarly, but for nearest-neighbours, this argument holds for the NTCDA lattice: electrons hopping from the marked dark molecule in Fig.\,\ref{fig: fig2}b to the lower-left bright molecule face on their right side the broad Kondo resonance of the dark molecule, while on the left side, they face the strong Kondo resonance of the bright molecule 
(see maps in Fig.\,\ref{fig: fig2}). For electrons in opposite directions, the situation is reversed.
This symmetry breaking does not occur for hopping between two bright or two dark molecules. 

With this model, we can reproduce our observations as the comparison of Fig.\,\ref{fig: fig4}a and \ref{fig: fig4}b shows. When the tip is placed above a bright molecule at site $i$ and $V=V^{0}_{\rm TP}$ (Fig.\,\ref{fig: fig4}c), the 
wavefunction $\Psi_0$ which crosses $E_{\rm F}$ is mainly localised at the site $i$ underneath the tip (Fig.\,\ref{fig: fig4}d). 
However, when the tip is placed 
between two bright molecules 
along the $\vec{a}_1$ direction, two transitions occur: At $V=V^{-}_{\rm TP}$ the antisymmetric superposition $\Psi_{-}\approx\sqrt{0.5}(\phi_{\text{A},i}-\phi_{\text{A},i+1})$ crosses $E_{\rm F}$ (Fig.\,\ref{fig: fig4}e), where $\phi_{\text{A},i}$ is the 
orbital wavefunction of the individual bright molecule at site $i$. This is then followed by the observed gap between the rings, and finally by the crossing of the symmetric superposition $\Psi_{+}\approx\sqrt{0.5}(\phi_{\text{A},i}+\phi_{\text{A},i+1})$ from $E_{\rm F}$ at $V=V^{+}_{\rm TP}$ (Fig.\,\ref{fig: fig4}f). Thus, at this tip position, each QPT involves a coherent superposition between two neighbours.

When we increase the bias further, more and more rings cross each other, and when the tip is placed on a dark molecule, as many as four lattice sites are coherently brought into the paramagnetic state at $V=2.22$\,V. This is remarkable because even though the gated molecules have a lateral distance of $0.95$\,nm from the tip apex, we only see a minor reduction of the TP intensity compared to the situation when the tip is on top of a bright molecule, clearly revealing the coherent lattice nature of the effect. Importantly, without the complex hybridisation between the sublattices, we cannot reproduce correctly the experimentally observed intensity modulation (see Supplementary Note 5). 

To summarise, taking advantage of sub-angstrom-resolution spectroscopic imaging offered by STM, we unraveled the role of strong electronic correlations in the formation of Hubbard bands and Kondo resonances in a self-assembled molecular  lattice. Build out of simple molecular building blocks, the lattice can reach macroscopic dimensions.  
We demonstrated that the ground state of the lattice can be controlled and manipulated by the local electric field of the STM tip, which drives the charge localised in the Hubbard bands across $E_{\rm F}$ and leads to a fundamental change of the lattice properties from a KS lattice to a PM metal.
We found clear signals of the quantum nature of this transition as finite fluctuations drive the transition even at zero temperature. 
For specific tip positions, interactions between neighbouring building blocks lead to a closed sequence of QPTs involving several molecules.
These sequences manifest themselves in ring-like structures which coherently interact with each other even nanometres away from the tip's position and create an interference pattern in real space. Finally, we showed that these interference patterns in the $dI/dV$ maps can be effectively simulated with a single-particle model Hamiltonian which has a complex hybridisation.

The observed QPTs shed light on the fragility of the correlations in supramolecular Kondo lattices. Even though the single-ion Kondo effect in our system has a high $T_{\rm K}$ originating from the strong interactions between the substrate and the SOMO, which would require a very large $\vec{E}$-field to be brought to $E_{\rm F}$, the lattice singlets are fragile objects that can be broken by relatively moderate fields. 

Our study not only expands the current understanding of strong correlations and topological properties in 2D materials but also offers a new supramolecular platform for further explorations. This can be done particularly by engineering complex structures out of molecular building blocks with the possibility of arbitrary pattering using molecular manipulation techniques and application of long-range gate operations at atomic scales.

\backmatter
\section*{Methods}
\subsubsection*{Experimental procedure}
The measurements were taken using a home-built combined STM and atomic force microscope operating in ultrahigh vacuum ($\leq 10^{-10}$\,mbar), at fields perpendicular to the sample surface of up to $B_z=14$\,T, and, if not mentioned otherwise, at a base temperature of $T=1.2-1.4$\,K. $dI/dV$ spectra were acquired by modulating the bias voltage with a sinusoidal $V_{m}$ of frequency $617$\,Hz and employing a lock-in amplifier. Note, the number of gated molecules is only limited by tip induced instabilities of the molecular lattice that start to occur at biases $V\gtrsim 2.4$\,V. 
The bare Pt tip was conditioned by indentation into the Ag(111) surface. The Ag(111) crystal was prepared by repeated cycles of Ar$^{+}$ sputtering and annealing. Afterwards, NTCDA molecules were deposited onto the clean Ag(111) surface held at room temperature from a Knudsen cell molecule evaporator.

\subsubsection*{Data analysis and fitting}
The statistical analysis of the $dI/dV$ grid spectra in Fig.\,\ref{fig: fig2} was performed by using \textit{feature-detection scanning tunnelling spectroscopy} (FD STS) \cite{Sabitova2018}. For that a second order Savitzky-Golay filter (SGF) with a sliding window of $\pm2\mbox{ mV}$ was applied at each data point. The SGF performs a quadratic fit $(y=a_{i}\,x^2+b_{i}\,x+c_{i})$, enabling the determination of peak positions when $b_{i}$ changes sign and $a_{i}>0$. Subsequently, a weight $w$ is assigned to each peak by determining the intensity difference $b_{i}-b_{i+1}$ between the neighbouring extrema in the first derivative of the signal, i.\,e.\ at $a_{i}=0$. This $w$ is then mapped producing the images in Fig.\,\ref{fig: fig2}b, and used as weight for the averaged spectra shown in Fig.\,\ref{fig: fig2}a.

The fits of Fig.\,\ref{fig: fig2}a are least-square fits of the Fano line-shape $g(eV)\propto g_0+(q+\epsilon)^2/(1+\epsilon^2)$, with $g$ as the differential conductance ($dI/dV$), $g_0$ as the background conductance, $\epsilon=(eV-\epsilon_0)/\Gamma$, $\epsilon_0$ as the center of the resonance, $\Gamma$ as its width, and $q$ as the Fano form-factor \cite{Fano1961}. Because $\Gamma\ll k_{\rm B}T$ and $\Gamma\ll eV_{m}$, we neglected temperature and modulation broadening effects.

The empirical Eq.\,\ref{eq:gamma} is motivated by the following assumptions: The quantum fluctuations are linearly dependent on the number of available states which can be reached when crossing the QPT, i.\,e., for the transition from the non-magnetic Kondo singlet state to the paramagnetic state it depends on the ratio between magnetic field energy $g\mu_{\rm B}\vec{B}$ and thermal energy $k_{\rm B}T$. This changes from $\overline{n}$=1 when the spin degrees are frozen out, to $\overline{n}$=2 when $\vec{B}=0$ resulting in a total broadening of $\overline{n}\,\Gamma_0'$. At $T>0$, an additional thermal broadening $\propto k_{\rm B}T$ occurs due to the Fermi-Dirac distribution of the tip's electrons which smears out any measurement in STM spectroscopy. Additionally, we have to account for the thermally activated broadening of the paramagnetic spin when $\overline{n}>1$. Both thermal broadening can be combined to $\propto k_{\rm B}T\sqrt{\overline{n}}$. Thus, for $\overline{n}=1$ only the tip effect is accounted for, while at $\overline{n}=2$, the two thermally activated broadening sources are geometrically added to the prefactor $\sqrt{2}$.
\subsubsection*{Tight-binding calculation} 
We used a single-particle tight-binding Hamiltonian:
\begin{linenomath*}
\begin{equation}
    \mathcal{H} = \mathcal{H}_{\rm sub} + \mathcal{H}_{\rm hyb}.
    \label{eq: model_Hamiltonian}
\end{equation}
\end{linenomath*}
Here, $H_{\rm sub}$ accounts for the on-site energies and the hopping within each sublattice (A$=$ bright, B$=$ dark molecules) and is given by:
\begin{linenomath*}
\begin{equation}
    \mathcal{H}_{\rm sub} = \sum_{\nu = \rm A,B}  \left[ \sum_{i} \left(E_{\nu}+U_{\nu,i}^{\rm eff} \right) \hat{c}_{\nu,i}^\dagger \hat{c}_{\nu,i}  - 
    \sum_{ \langle\langle i,j \rangle\rangle}  t_{\nu,ij}\left[\hat{c}_{\nu,i}^\dagger \hat{c}_{\nu,j} + h.c.\right] \right],
\end{equation}
\end{linenomath*}
and $\mathcal{H}_{\rm hyb}$ for the hybridisation between the two sublattices:
\begin{linenomath*}
    \begin{equation}
    \label{eq: Hyb}
        \mathcal{H}_{\rm hyb} = -  \sum_{\langle i,j \rangle} \left[V_{\rm AB}\,e^{i\theta_{ij}}\,\hat{c}_{\text{A},i}^{\dagger}\hat{c}_{\text{B},j}  + h.c.\right].
    \end{equation}
\end{linenomath*}
The creation $\hat{c}^{\dagger}_{\rm \nu,i}$ and annihilation operators $\hat{c}_{\rm \nu,i}$ act on the quasiparticle at site $i$ of the sublattice $\nu$. 
The single bracket $\langle i,j \rangle$ and double bracket $\langle\langle i,j \rangle\rangle$ in the sums stand for the nearest- and next-nearest-neighbour sites, where we allowed the hopping amplitude $t_{\nu,ij}$ to differ in each sub-lattice ($\nu$) and also along the short and long unit vectors. We used the energies of the LHB and UHB for the on-site energies $E_{\rm A}$ and $E_{\rm B}$. However, the simulations remain indifferent to the exact value of $E_{\rm B}$ as long as $E_{\rm B}>0$. 
To include the local $\vec{E}$-field of the tip we added at each lattice site $U_{\nu,i}^{\rm eff}=\alpha  z\,\vec{e}_{z}\cdot\vec{E}(d_{\nu,i})$, where $d_{\nu,i}$ is the lateral distance between the tip apex and site $i$ at sublattice $\nu$. The $dI/dV$ maps of Fig.\,\ref{fig: fig4}b were then calculated by diagonalization of Eq.\,\ref{eq: model_Hamiltonian} at each pixel and by taking the absolute square of the coherent sum of all spatially extended wavefunctions $\Psi_n$ which energy $\xi_n$ lie within the derivative of Fermi-Dirac distribution
%
\begin{linenomath*}
\begin{equation}
\label{eq: didv}
    dI/dV \propto  \sum_{n} \left \lvert \Psi_n \right\rvert^2 \left . \frac{\partial f_{\rm F}(\varepsilon)}{\partial \varepsilon}  \right \vert_{\varepsilon = \xi_{n}} .
\end{equation}
\end{linenomath*}
%
Note, that each wavefuction $\Psi=\sum_{\nu=\text{A,B}}\sum_i a_{\nu,i}\phi_{\nu,i}$ is a sum of spatially extending Warnier-like orbitals $\phi_{\nu,i}$ which are centred at the molecules at site $(\nu,i)$, with complex coefficients $a_{\nu,i}$. We used $s$-like orbitals for the representation in Fig.\,\ref{fig: fig4}d-f, whereby we mark that  the shape of the $\phi_{\nu,i}$ did not play a role in the calculation of the $dI/dV$.
\bmhead{Acknowledgments}
We thank Christian Ast,  Johann Kroha, Christopher Leon, and 
Nicol\'as Lorente for fruitful discussions. M.T. acknowledges support from the Heisenberg Program (Grant No. TE 833/2-1) and the Priority Program (SPP 2244) of the German Research Foundation.
\bmhead{Author contributions}
R.T., F.S.T., and M.T. conceived the experiment.
S.A., T.E., R.T., A.S., and Y.W. performed the STM measurements. S.A., M.T., T.E., and R.T. analysed the experimental data. S.A. and M.T. developed the theoretical modelling and tight-binding calculations. S.A. and M.T. wrote the manuscript with significant input from T.E., H.L., C.W., F.S.T., and R.T. All authors discussed the results and reviewed the manuscript.

\bibliographystyle{naturemag}
\bibliography{sn-bibliography}

\end{document}